\documentclass[12pt,a4paper]{article}
\usepackage{amsmath,amssymb,amsfonts,booktabs,colortbl}
\usepackage{graphicx}

\newcommand{\nn}{\nonumber} 
\newcommand{\be}{\begin{equation}} 
\newcommand{\ee}{\end{equation}} 
\newcommand{\bea}{\begin{eqnarray}}  
\newcommand{\eea}{\end{eqnarray}}

\newcommand{\hc}{\mathrm{h.c.}}

\newcommand{\SM}{\mathrm{SM}}

\newcommand{\tinymath}[1]{{\tiny{\mbox{$#1$}}}}
\newcommand{\scriptmath}[1]{{\scriptsize{\mbox{$#1$}}}}

\newcommand{\ctoprule}{\toprule[0.4mm]}
\newcommand{\cbottomrule}{\bottomrule[0.4mm]}

\begin{document}


\begin{flushright}
UG-FT-262/09 \\
CAFPE-132/09 \\
UCRHEP-T479\\

November 17, 2009
\end{flushright}
\vspace*{5mm}
\begin{center}

\renewcommand{\thefootnote}{\fnsymbol{footnote}}

{\Large {\bf Evidence for right-handed neutrinos at a neutrino factory  
}} \\
\vspace*{1cm}
{\bf F.\ del Aguila$^a$}\footnote{E-mail: faguila@ugr.es},
{\bf J.\ de Blas$^a$}\footnote{E-mail: deblasm@ugr.es}, 
{\bf R.\ Szafron$^c$}\footnote{E-mail: szafron@us.edu.pl},
{\bf J.\ Wudka$^b$}\footnote{E-mail: jose.wudka@ucr.edu} \\
and
{\bf M.\ Zra{\l}ek$^c$}\footnote{E-mail: zralek@us.edu.pl}

\vspace{0.25cm}

$^a$ Departamento de F\'{\i}sica Te\'orica y del Cosmos and CAFPE,\\
Universidad de Granada, E-18071 Granada, Spain \\
\vspace{0.25cm}
$^b$ Department of Physics and Astronomy,\\
University of California, Riverside
CA 92521-0413, USA \\
\vspace{0.25cm}
$^c$ Department of Field Theory and Particle Physics, Institute of Physics, \\
University of Silesia, Uniwersytecka 4, 40-007 Katowice, Poland \\

\end{center}
\vspace{.5cm}

\begin{abstract}
 
\noindent 
We emphasize that a muon based neutrino factory could show the 
existence of light right-handed neutrinos, if a deficit in the 
number of detected events is observed at a near detector. 
This could be as large as $\sim 10 \% $ if the size of the new 
interactions saturates the present limits from electroweak 
precision data, what is not excluded by the oscillation 
experiments performed up to now. 
A simple model realizing such a scenario can be obtained 
adding right-handed neutrinos to the minimal Standard Model, 
together with an extra scalar doublet and a triplet of 
hypercharge 1. In this case, however, the possible 
deficit is reduced by a factor of $\sim 3$, and the Yukawa couplings 
must be adequately chosen. 
This is also generically required if lepton flavour violation 
must be below present bounds.
\end{abstract}

\renewcommand{\thefootnote}{\arabic{footnote}}
\setcounter{footnote}{0}


\section{Introduction} 
Neutrino oscillations provide the only observational evidence 
of new physics beyond the minimal Standard Model (SM). 
At present these oscillations can be fully explained by introducing 
neutrino masses and the corresponding charged current mixing \cite{Amsler:2008zzb}. 
These observations, however, cannot distinguish between  
Dirac or Majorana neutrinos, 
nor they require new interactions (NI)~\footnote{In what follows we will 
ignore the LSND data \cite{Aguilar:2001ty}.}.
The obvious question is then, where do we have to look 
in order to determine the neutrino character and/or 
to observe possible NI involving light neutrinos? 
This will become especially relevant when we face the 
need to interpret new data with higher statistics and
precision, as foreseen from a neutrino factory 
\cite{Abe:2007bi}. 
In the following we show that present experimental 
constraints leave room, corresponding to a $\sim 10 \% $ deficit  
in the expected number of events in appropriate processes, 
for observing NI involving 
light right-handed (RH) neutrinos. Such a scenario 
can be easily realised with a mild extension of the SM,  
through the addition of additional scalar
weak isodoublet (to be denoted by $\eta$), 
and a scalar iso-triplet of hypercharge $1$,  
(denoted by $\Delta$), besides three RH neutrinos, $\nu _{R}^i$. 
But it requires an adequate choice of Yukawa couplings 
to suppress lepton flavour (LF) violation; moreover, 
for this particular model, the 
deficit allowed by current electroweak precision data (EWPD) 
is reduced by a factor of $\sim 3$ compared 
to the general case above where arbitrary NI are parameterized by 
gauge-invariant dimension six operators with unrelated coefficients.  
 
Within the SM muons only decay into left-handed (LH) neutrinos. 
Even if the spectrum is enlarged to include their RH counterparts, 
these are not produced in such decays because they have no gauge interactions, 
and neutrino masses are negligible. 
On the other hand, if other 
interactions are present in nature, 
a muon based neutrino factory could inject an 
admixture of neutrinos with both chiralities. 
We will show below that the limits on NI involving 
RH neutrinos are to a large extent those derived 
from (inverse) muon decay, and therefore relatively weak. 
This then is a promising reaction where to 
look for new physics effects in the neutrino system.

Let us first, however, state our setup. 
We assume that the three light neutrinos are 
Dirac-type neutrinos, i.e. that there are three light neutrino singlets 
beyond the minimal SM, and that lepton number (LN) is conserved. 
In practice this is not a restriction  
on the light neutrino character, but on the type of NI.  
Notice that neutrino masses are negligible in all experiments 
performed up to now, except in neutrino oscillations (and eventually in 
neutrino-less double $\beta$ decay, $0 \nu \beta \beta $). 
Thus, we can assume that all interactions conserve LN because 
neutrino masses are much smaller than the energy relevant in 
the processes considered and/or the experimental precision is much 
lower than the size of the effects proportional to them,  
as it is the case for all foreseen experiments not involving neutrino 
oscillations and excluding $0 \nu \beta \beta $. 
Though we could consider LN violating NI, 
their effects can be ignored in our analysis  
\footnote{An effective theory only involving the light SM fields 
and invariant under the SM gauge group has only one dimension 
five operator violating LN \cite{Buchmuller:1985jz}, 
the famous Weinberg operator \cite{Weinberg:1979sa} 
giving Majorana masses to the LH neutrinos and then negligibly small. 
There is no operator violating baryon minus lepton (B$-$L) 
number of dimension six \cite{de Gouvea:2007xp}. 
Thus, operators of dimension six violating LN also violate baryon 
number (BN), and then involve quarks and will play no r\^ole in our 
analysis. If the effective theory also includes RH neutrinos, as in our 
case, there are two additional dimension 5 LN violating operators
\cite{delAguila:2008ir}, one of them generates a magnetic coupling
for the RH neutrinos and is very strongly constrained when the
$ \nu_R$ are light \cite{Aparici:2009fh}, the other generates
a correction to the $\nu_R$ Majorana mass term and is
therefore also negligible. There is only one dimension six operator violating B$-$L 
(and LN) but involves four RH neutrinos \cite{delAguila:2008ir}, 
and then is uninteresting for us. 
The other dimension six operators violating LN also violate 
BN, and can be also ignored.}, 
as can be the LN violating effects from neutrino masses.

Hence, the NI effects we are interested in will be relatively large and 
LN conserving; whereas neutrino masses will be safely taken 
to vanish. The effective Lagrangian describing such a scenario 
will not distinguish between ({\it i.e.} approximates equally well) the case 
of exact LN conservation with very small Dirac masses 
for the three light neutrinos, and the case 
of negligible Majorana masses 
for the six light neutral fermions, 
the only vestige of the very slightly broken LN in this case. 
This was explicitly proved in \cite{Langacker:1988cm} 
for (inverse) muon decay assuming no additional constraints on NI. 
Note that by similar arguments we can also neglect LF 
violation induced by light neutrino mixing 
(to a very good approximation). 

The most general four-fermion effective Hamiltonian 
describing muon decay reads  
\be
\begin{split}
{\cal H}_{\mu\rightarrow \nu \overline{\nu}e}&=\frac{4 G_F}{\sqrt{2}} 
\sum_{\tinymath
{\begin{array}{c}\alpha, \beta=L,R\\\gamma=S,V,T\end{array}}} 
g_{\alpha\beta}^\gamma \left(\overline{e}\ \Gamma^\gamma\ \nu _\alpha ^e\right)
\left(\overline{\nu ^\mu _\beta}\ \Gamma^\gamma\  \mu\right) + \hc \ ,
\end{split}
\label{mudecH}
\ee
where $\alpha , \beta $ label the chirality of the neutrinos, while
$ \gamma $ refers to the 
Lorentz character of the interaction (scalar, vector and tensor).  
The present limits on the size of the various coefficients 
will be discussed in the following section, here we
merely remark that the  two couplings 
$g^V_{LL},~ g^S_{RR}$ are also associated with the largest departure 
from the SM predictions for the number of events to be detected by a 
neutrino factory.

There are many other available electroweak precision data 
that can be used to indirectly constrain the Hamiltonian
(\ref{mudecH}), 
and in particular those two couplings. 
To derive such restrictions one 
can take two routes. The first one consists in 
re-writing (\ref{mudecH}) as a linear combination of 
higher-dimensional operators
invariant under $SU\left(3\right)_c\times SU\left(2\right)_L\times U\left(1\right)_Y$,
and adding these operators to the SM Lagrangian 
\cite{Gavela:2008ra}; the coefficients
of the resulting effective theory can then be bound using experimental data
at all available energies. In this approach the physics responsible for
generating these operators is left unspecified, except for the
requirement that its characteristic scale lie beyond the electroweak
scale. The advantage of this approach is its generality, since it is not
 tied to any specific assumption about the physics beyond the SM,
its disadvantage is the proliferation of coefficients, and
the fact that more than one gauge-invariant operator contributes to each
term in (\ref{mudecH}). 

The second route is to extend the SM by adding a specific set 
of new fields (such as the $ \eta$ and $\Delta$ mentioend previously)
and interactions. This has the advantage of providing
a specific scenario for the physics beyond the SM, but the results
obtained are often specific to the assumptions made in constructing
the model. At scales below those of the heavy particles this
model will reduce to an effective theory of the type mentioned
above, except that the effective-operator coefficients are all
expressed in terms of a small number of parameters and can be
constrained more tightly.

In the following we will examine both of these possibilities. In the next
section we consider
effective Lagrangian approach, which we denote by E1, where we choose
a set of 4-fermion effective operators that generate  (\ref{mudecH})
at low energies and have little impact on other electroweak observables. 
In section \ref{sec:model} we will also consider a specific 
extension of the SM, which we denote by E2, based on an extended scalar sector. 
Finally, in the last two sections we discuss the implications of these SM 
extensions for neutrino oscillations and other experiments, respectively. 

\section{Electroweak precision data constraints}  
\label{EWPD} 
One effective-Lagrangian extension of the SM, which denote by E1, consists
in adding to the SM the following set of effective operators
\begin{eqnarray}
& & \!\!\!\!\!\!\!\!\! \frac{4 G_F}{\sqrt{2}} \;
\left[\, g_{LL}^S \left(\overline{e_R}l_L^e\right)
\left(\overline{l_L^\mu}\mu_R\right) 
+g_{RR}^S\left(\overline{l^e_L}\nu_R^e\right)
\left(\overline{\nu_R^\mu}l^\mu_L\right) \right. \nonumber \\
& + & g_{LR}^S\left(\overline{e_R}l_L^e\right)i\sigma_2 
\left(\overline{\nu_R^\mu}l^\mu_L\right)  
-g_{RL}^S \left(\overline{l_L^e}\nu_R^e\right)i\sigma_2
\left(\overline{l_L^\mu}\mu_R\right) \nonumber \\
& + & \frac{\delta g_{LL}^V}{2}\left(\overline{l^e_L}\gamma^\mu l_L^\mu\right)
\left(\overline{l_L^\mu}\gamma_\mu l^e_L\right)
+g_{RR}^V\left(\overline{e_R}\gamma^\mu \nu_R^e\right)
\left(\overline{\nu^\mu_R}\gamma_\mu \mu_R\right)
\label{mudecHE1}\\
& - & \frac{g_{LR}^V}{v^2} \left(\overline{l^e_L}
\sigma_a\gamma^\mu l_L^e\right)\left(\phi^Ti\sigma_2 \sigma_a \phi\right)
\left(\overline{\nu^\mu_R}\gamma_\mu \mu_R\right) \nonumber \\
& + & \frac{g_{RL}^V}{v^2}\left(\overline{e_R}\gamma_\mu \nu^e_R\right)
\left(\phi^\dagger \sigma_a i\sigma_2\phi^*\right)
\left(\overline{l^\mu_L}\sigma_a\gamma^\mu l_L^\mu\right)
\nonumber \\
& + & g_{LR}^T \left.\left(\overline{e_R}\sigma^{\mu\nu} l_L^e\right)i\sigma_2
\left(\overline{\nu_R^\mu}\sigma^{\mu\nu} l^\mu_L\right)
-g_{RL}^T\left(\overline{l^e_L}\sigma^{\mu\nu} \nu_R^e\right)i\sigma_2
\left(\overline{l_L^\mu}\sigma^{\mu\nu} \mu_R\right)\right]+\hc\ , \nonumber
\end{eqnarray}
where $l_L^f$ stands for one of the three SM lepton doublets
($ f = e,~ \mu,~\tau$) and $\phi$ 
is the SM Higgs doublet, and $v \simeq 246$ GeV its vacuum expectation value. 
We do not use the basis proposed in \cite{Buchmuller:1985jz} 
for writing the beyond the SM dimension six operators; still it
is important to note that this basis must be extended 
to include the light RH neutrinos \cite{delAguila:2008ir} 
\footnote{Although in that reference the RH neutrinos are 
assumed to have masses of few hundreds of GeV. 
(Note that the operator $ {\cal O}_{QNdQ} $ 
in Eq. (7) of that reference is redundant.)}. 
E1 is manifestly invariant under 
$SU\left(3\right)_c\times SU\left(2\right)_L\times U\left(1\right)_Y$, 
and LF (and LN) conserving. 
Note that $g_{LR}^V$ and $g_{RL}^V$ are associated with operators of dimension eight: 
$\overline{\nu^\mu_R} \gamma^\mu \mu_R$ and $\overline{e_R} \gamma^\mu \nu_R^e$ 
have hypercharges $Y=-1$ and $1$, respectively, and we cannot construct 
a gauge invariant vector with opposite hypercharge  using only
$\overline{l_L}$ and $l_L$; at low energies (\ref{mudecHE1}) reduces to
(\ref{mudecH}) with the definition $ g^V_{LL} \equiv 1 + \delta g^V_{LL}$.

In (\ref{mudecHE1}) we have chosen to write all dimensional coefficients in terms
of $v$. This implies that the natural size for the coefficients is
\be
g^\gamma_{\alpha \beta} \sim
\left\{
\begin{array}{ll}
(v/\Lambda)^4 & \hbox{for}~ g^V_{LR,\,RL}\ , \cr
(v/\Lambda)^2 & \hbox{otherwise}\ ,
\end{array} \right.
\label{NPlimits}
\ee
where $ \Lambda $ denotes the heavy scale of the physics responsible for generating
the corresponding operator.

A characteristic feature of this particular extension is that, except for the effect 
on the muon decay constant and inverse muon decay, 
EWPD are 
blind to the operators in Eq.~(\ref{mudecHE1}).  
For instance, although $g^S_{LR}$ and $g^T_{LR}$ 
contribute to $e^+e^- \rightarrow \overline{\nu_\mu}\nu_\mu$
and affect the $Z^0$ invisible width, the effects is
negligible compared to the $Z^0$ pole contribution.  
E1 does not contribute to LF violating processes either, because 
it does not include LF violating operators. 
Similarly, universality is preserved since the gauge couplings 
to leptons remain the same as in the SM.

It must be emphasized, however, that in specific models 
(such as E2, described in section \ref{sec:model} below)
the various coefficients in Eq. (\ref{mudecH}) are written in terms
of the fundamental parameters of the theory, these parameters
appear in all other interactions, 
and in general will contribute to other 
observables, which in turn can be used to stringently constrain 
$g_{\alpha\beta}^\gamma$. In fact, it is non-trivial 
to find models where these limits are not so strict they exclude further 
observable effects from (\ref{mudecH}).

\begin{figure}[ht]
\begin{center}
\includegraphics[width=9cm,height=6cm]{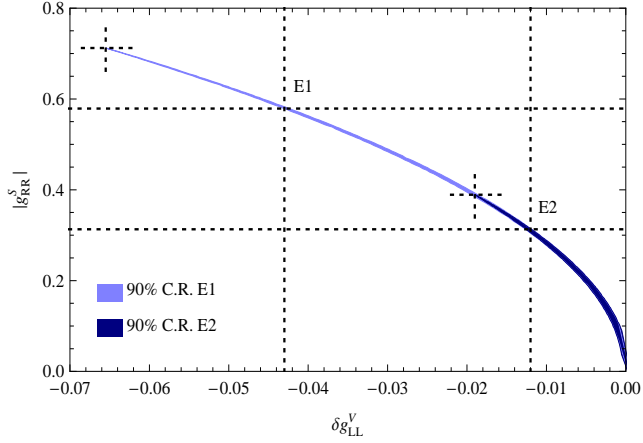}
\caption{90\% C.L. bounds for E1 case $(b)$ in 
Eq. (\ref{E1limits}); and the same for the SM extension 
E2 in next section. The narrow bands  between the
origin and the crosses define the 90\% confidence 
region for the global fit to the two parameters for
 E1 (left cross) and E2 (right cross) respectively.}
\label{90CRannul}
\end{center}
\end{figure}

As we argue below the largest departure from the SM 
predictions is parameterised by $g^V_{LL}$ and $g^S_{RR}$. 
Constraints on these two 
operators~\footnote{See \cite{Biggio:2009nt} 
and reference therein for a fit to 
operators involving only left-handed neutrinos.} 
$\left(\overline{l^e_L}\nu_R^e\right)
\left(\overline{\nu_R^\mu}l^\mu_L\right)$ and 
$\left(\overline{l^e_L}\gamma^\mu l_L^\mu\right)
\left(\overline{l_L^\mu}\gamma_\mu l^e_L\right)$ 
in Eq. (\ref{mudecHE1})
can be obtained using current data, as described in the Appendix.
At 90\% C.L. we
find
\be
\begin{array}{ccc}
{\rm case} & \left|g^V_{LL} \equiv1 + \delta g^V_{LL}\right|& |g^S_{RR}| \cr
\vspace{-3mm} & & \cr 
(a) & >0.960 & <0.550 \cr
(b) & >0.957 & <0.579 \cr
(c) & >0.9998 & <0.054 \cr
\end{array}
\label{E1limits}
\ee
where the limits in case $(a)$ are obtained directly from
(\ref{mudecH}) using muon decay and
$ \nu e \to \nu \mu $ data \cite{Amsler:2008zzb,Fetscher:1986uj};
in case $(b)$ from a global fit using precision data to the 
operator coefficients with the SM parameters fixed at their 
best-fit values~\footnote{These limits stay mainly unchanged if the SM parameters 
are also left free in the global fit.};
in case $(c)$ as in case $(b)$ but taking one effective-operator
coefficient to be non-zero at a time. This last possibility,
though often adopted for simplicity is seldom realistic:
as noted previously one must expect the heavy physics to generate
several operators with related coefficients, so that
fits allowing for several non-zero interactions 
become compulsory. 
Even more, if the new physics is to have sizable indirect
effects, then it must also {\em conspire} to preserve the
excellent fit of the SM to the EWPD at the 0.1\%\ level 
\cite{Amsler:2008zzb}. 
In Fig. \ref{90CRannul} we plot the bounds for case $(b)$. 
The results in (\ref{E1limits}) indicate that the
two parameters in the fit are highly 
correlated. This is expected since 
the fit is dominated by the constraint on the 
strength of the muon decay constant $G_\mu$, which is 
proportional to the SM Fermi constant $G_F$ in 
Eq. (\ref{mudecH}):
\bea
\label{GmuGF}
G_\mu&=&G_F\left[\left|1 + \delta g_{LL}^V\right|^2
+ \left|g_{RR}^V\right|^2
+ \left|g_{LR}^V\right|^2
+ \left|g_{RL}^V\right|^2 \right. \nn \\
&+&\frac 14 \left( \left|g_{LL}^S\right|^2 
+ \left|g_{RR}^S\right|^2 
+ \left|g_{LR}^S\right|^2
+ \left|g_{RL}^S\right|^2 \right)  \\
&+&\left.4\left(\left|g_{RL}^T\right|^2+
\left|g_{LR}^T\right|^2\right)\right]^{\frac 12}\equiv G_F A\ ,\nn
\eea
with the proportionality constant $A$ restricted by the 
global fit to the interval
\be
0.9997<A<1.0004~\mbox{ at } 90\%\mbox{ C.L.}\ .
\label{Aconstr}
\ee
This bound alone constrains the coefficients to a narrow  band in 
the $\delta g^V_{LL} - \left|g^S_{RR}\right|$ plane 
\be
|g_{RR}^S|^2  \simeq - 8\ \delta g_{LL}^V\ ,
\label{correlation} 
\ee
as depicted in Fig. \ref{90CRannul}. 

Using the bounds above we can derive limits on the scale
of new physics responsible for the two operators
being considered. Using (\ref{NPlimits})
and assuming the underlying physics is weakly coupled we find
the weak constraints $ \Lambda > 130 $ GeV for $ g^S_{RR}$
and $ \Lambda > 500 $ GeV for $ \delta g^V_{LL} $.

As stressed in the introduction, 
we are interested in those interactions in the muon decay
effective Hamiltonian that could allow for the largest deviation from the  
SM prediction of the number of events that may be eventually detected 
at a neutrino factory. In general, while it is clear that  
a negative
$\delta g_{LL}^V$ is strongly favored
for this to be possible (see Eqs. (\ref{GmuGF}, \ref{Aconstr})), 
one may wonder whether any  
of the other interactions
involving a RH neutrino may play the role of $g_{RR}^S$. This could be  
the case for $g_{RR}^V$ or the
LR and RL operators in Eq.~(\ref{mudecHE1}). However, unlike for  
$g_{RR}^S$, which cannot be separated from $\delta g_{LL}^V$ in muon  
decay experiments since we do not measure the polarization of the  
final neutrinos, all the other interactions are contrained by  
the absence of any significant deviation from the V$-$A prediction in  
the spectrum of the outgoing electrons. In particular, $\left|g_{RR}^V 
\right|<0.034$. On the other hand, 
the bounds for some of the LR and RL interactions are  
relatively weak, but they also generate radiative corrections
to the neutrino masses and, assuming naturality, 
are constrained by the associated
upper limits~\cite{Prezeau:2004md}: $\left|g_{LR}^S\right|,\left|g_{RL}^V 
\right|,\left|g_{LR}^T\right|<10^{-2}$ and $\left|g_{RL}^S\right|, 
\left|g_{LR}^V\right|,\left|g_{RL}^T\right|<10^{-4}$.
Therefore, when we consider the data from inverse muon decay, where  
the incident neutrinos come from pion decays and then are LH, and  
derive a relatively weak bound on $\delta g_{LL}^V$, 
this can be only compensated by  
$g_{RR}^S$ in order to satisfy (\ref{Aconstr}),  but
in contrast with the remaining operators in 
Eq. (\ref{mudecHE1}) has no further constraints.

\section{A simple SM extension} 
\label{sec:model}
Let us discuss a simple model realising the former scenario. 
It extends the SM including besides three RH neutrinos $\nu ^i_R$ 
with zero LN, 
a second scalar doublet $\eta$ 
with LN equal to $-1$, and a scalar triplet $\Delta$ 
with hypercharge $1$ and LN equal to $-2$, 
\begin{equation}
\eta = \Big(\begin{array}{c}\eta ^+\cr \eta ^0 \end{array}\Big) \; ,  \;\;\;  
\Delta = \Big(\begin{array}{cc} \Delta ^+ &\sqrt 2 \Delta ^{++} \cr 
\sqrt 2 \Delta ^0 & -\Delta ^+ \end{array}\Big) \ , 
\end{equation}
as in \cite{Ma:2000cc} and \cite{Abada:2007ux}, 
respectively. They both can  acquire a vacuum expectation 
value through very small LN violating couplings to the SM Higgs 
(which carries zero LN), and provide masses and mixings to the light neutrinos,
which must be then fit the observed spectrum in oscillation 
experiments 
\footnote{There can be also tiny RH neutrino mass terms.}.    
But, these couplings, as the neutrino masses, 
can be safely neglected in our analysis:
we can assume that light neutrinos are massless, 
and that LN and LF are both conserved. 
Thus, besides the kinetic term for the RH neutrinos, 
the SM fermionic Lagrangian acquires only two more terms 
\be
-f_{ij} \overline{\nu_{R}^i} l^{j T} i\sigma_2\eta 
-\lambda_{ij}l_{L}^{i T} Ci\sigma_2 \Delta l_{L}^j +\hc\ , 
\label{addition}
\ee
with only $f_{ee, \mu \mu}$ and $\lambda_{e \mu} = \lambda_{\mu e}$ 
potentially large enough to produce measurable effects in current
and near-future experiments. The integration of the extra scalars out 
results, in particular, in the following contributions to the 
muon decay effective Hamiltonian:
\be
\frac{4 G_F}{\sqrt{2}}
\left[\ g_{RR}^S\left(\overline{l^e_L}\nu_R^e\right)
\left(\overline{\nu_R^\mu}l^\mu_L\right)
+\frac{\delta g_{LL}^V}{2} \left(\overline{l^e_L}\gamma^\mu l_L^\mu\right)
\left(\overline{l_L^\mu}\gamma_\mu l^e_L\right)\ \right]+\hc\ , 
\label{mudecHE2}
\ee
with coefficients
\be
g_{RR}^S=-\frac{1}{2\sqrt{2}G_F}
\frac{f^\dagger_{ee}f_{\mu\mu}}{M_\eta^2} 
\quad \mbox{and} \quad 
\delta g_{LL}^V=-\frac{1}{\sqrt{2}G_F}
\frac{\lambda^\dagger_{\mu e}\lambda_{\mu e}}{M_\Delta^2}\ , 
\ee
respectively, where $M_{\eta , \Delta}$ stand for the scalar masses. 
The full set of dimension six operators arising from the integration of 
the scalar triplet $\Delta$ is given in \cite{Abada:2007ux}.

In this case, named E2 in the former section, the EWPD 
analysis presented in the Appendix implies more stringent bounds on 
$\delta g^V_{LL}$ and $g^S_{RR}$ than for E1: 
\be
\left|1 + \delta g^V_{LL}\right|>0.988 , \ 
{\rm and} \ 
|g^S_{RR}| < 0.313\ . 
\label{E2limits} 
\ee
The corresponding band is plotted in Fig. \ref{90CRannul}. 
The bounds obtained are tighter because, as emphasized previously, 
the integration of definite 
new physics also gives, in general,  operators contributing to 
other processes, further restricting the model. 
In the present case the integration of the $\Delta $
generates also the operator   
$\left(\overline{l^e_L}\gamma^\mu l_L^e\right)
\left(\overline{l_L^\mu}\gamma_\mu l^\mu _L\right)$, 
which has the same coefficient $\delta g^V_{LL}$ as 
$\left(\overline{l^e_L}\gamma^\mu l_L^\mu\right)
\left(\overline{l_L^\mu}\gamma_\mu l^e_L\right)$, 
and contributes to 
$\nu _\mu e \rightarrow \nu _\mu e$,
further restricting the allowed deviation from the SM predictions 
\footnote{In our approximation (implying negligible LN violation 
in the scalar sector) there are no tree-level contributions to the 
oblique parameters.}. 

LF violation is below experimental bounds because 
similarly to the E1 case the only 
non-negligible couplings in Eq. (\ref{addition}) are 
$f_{ee, \mu \mu}$ and $\lambda_{e \mu} = \lambda_{\mu e}$,
and because the scalar doublets and triplet mix very little, 
as required by approximate LN conservation. 
The absence of new $\tau$ couplings and that the SM 
gauge couplings stay unchanged guarantee the agreement 
with universality constraints on the lepton sector.

\section{Neutrino factory predictions} 
\begin{figure}[!]
\begin{center}
\includegraphics[width=9cm,height=6cm]{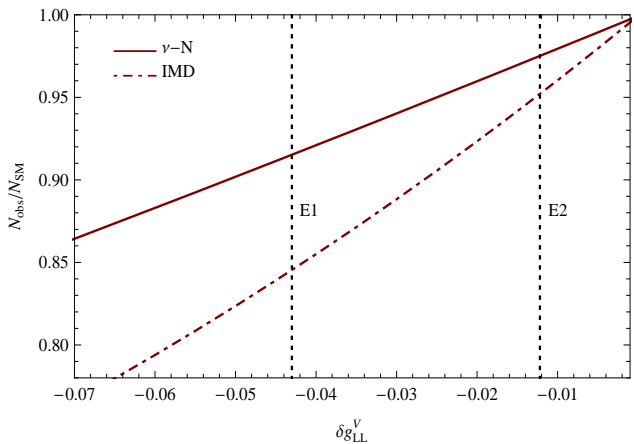}
\caption{Percentage of detected events in a near detector at a 
neutrino factory compared to the predicted number by the SM, 
as a function of the NI strength $\delta g_{LL}^V$. 
The solid (dotted-dashed) curve corresponds to neutrino-nucleon 
(inverse muon decay) collisions. The vertical lines stand for 
the E1 and E2 limits on $\delta g_{LL}^V$ 
in the text.}
\label{Defmuplot}
\end{center}
\end{figure}

The relevant phenomenological question is where could
the RH neutrinos be eventually observed if 
the $\delta g_{LL}^V$ and $g_{RR}^S$ 
four-fermion interactions are non-zero. 
Obviously, they can be probed in a more 
precise inverse muon decay experiment: 
a more precise measurement of this process could 
give evidence for those NI (or reduce the allowed deviation from the SM 
in Eqs. (\ref{E1limits},\ref{E2limits}) and Fig. \ref{90CRannul}). 
But a muon-based neutrino factory will also be sensitive to them. 
Indeed, if a substantial amount
of the neutrinos produced in muon decay are RH, 
a near-detector sensitive to neutrino-hadron collisions 
will observe a deficit in the same proportion, and this 
deficit would be twice as large if the detector could also measure 
the inverse muon decay process.   
In Table \ref{tab:Maxdef} we give the maximum deficits expected 
in the case that the new interactions saturate the $90\%$ C.L. bounds 
obtained in the previous sections. 
\begin{table}[t]
\begin{center}
\begin{tabular}{ c c c } \hline
\ctoprule
Process &E1& E2\\
\midrule
$\nu$-$N$&$\ \ 8.5\%\ (5.0\%)$&$2.5\%\ (2.5\%)$\\
IMD&$15.4\%\ (9.3\%)$&$4.8\%\ (4.8\%)$\\
\cbottomrule
\end{tabular}
\caption{Maximum deficit in the number of observed events in a near detector 
sensitive to neutrino-nucleon collisions ($\nu$-$N$) and inverse muon decay 
(IMD) for the two SM extensions discussed in the text. 
In parentheses we show the deficits expected in the case that the precision 
on the measurement of inverse muon decay is improved by a factor of 2.} 
\label{tab:Maxdef}
\end{center}
\end{table}
Whereas in Fig. \ref{Defmuplot} we show the predicted deficit as a function 
of $\delta g_{LL}^V$ for the SM completions considered. 

If the precision 
in the measurement of the inverse muon decay is improved by a factor of 2 
without modifying the central value,  
the limits on the NI would be strengthened and 
the allowed deficit of observed events at a
neutrino factory reduced; 
the corresponding percentages in such a case are given in parentheses 
in Table \ref{tab:Maxdef}. For the E1 extension 
the deficit would be reduced to $\sim 5 \%$, being the 90\% C.L. 
bounds in this case $\left|1 + \delta g^V_{LL}\right|>0.975$ and
$|g^S_{RR}| < 0.442$. In the E2 case the improvement in the 
precision of inverse muon decay would have no appreciable effect because 
this constraint would be weaker than the
one derived from $ e \nu $ elastic scattering.

For a neutrino factory to be sensitive to a deficit
$< 3\%$ (the smallest value
listed in Table \ref{tab:Maxdef}), 
the neutrino flux must be known with 
enough precision. Besides, large fluxes are also required to have a large 
number of events at a near detector in order to keep the statistical error 
small. 
For a detailed study about future neutrino factories see~\cite{Abe:2007bi}.  
Assuming $10^{21}$ muon decays in one year of operation, 
the number of expected $\nu$-$N$ events
at a near detector (such as the one described in
Table 1 of~\cite{Abe:2007bi}) 
is of the order of $10^9$; while the number of IMD events  
range from $10^4$ to $10^6$, depending on the energy and polarization  
of the decaying muon. 
Then, deficits even smaller than few per mille 
due to the injection of RH neutrinos 
could be eventually testable for such a large statistics. 
However, per mille deficits may be too small because the highest 
achievable precision in the determination of the flux is expected 
to be at most of $\sim 0.1 \% $. But being conservative, it could be 
up to a factor ten times larger, and then of the same order as the 
largest possible deficit for the E2 extension. 

At this point one may wonder about the consistency 
of these large deficits with the interpretation of 
present neutrino oscillation experiments summarized in Table~\ref{tab:NuOscExp}.  
\begin{table}[h]
\begin{center}
\begin{tabular}{ c c c c} \hline
\ctoprule
$\nu$ source & Experiment & Detection &\\
\midrule
Reactor&Palo Verde~\cite{Boehm:2001ik}, CHOOZ~\cite{Apollonio:2002gd}&$\overline{\nu_e} \not \rightarrow  \overline{\nu_\mu}$\\
($\beta$ decay)&KamLAND~\cite{:2008ee}&$\overline{\nu_e}\rightarrow\overline{\nu_x}$\\
\midrule
Solar&SNO~\cite{Aharmim:2005gt}, Borexino~\cite{Arpesella:2007xf}&$\nu_e \rightarrow  \nu_\mu$\\
\midrule
Atmospheric&Super Kamiokande~\cite{Ashie:2005ik}&$\nu_\mu\rightarrow \nu_\tau$&\\
($\pi$ and $\mu$ decays)& & &\\
\midrule
Accelerator&K2K~\cite{Ahn:2006zza}, MINOS~\cite{Michael:2006rx},&$\nu_\mu\rightarrow \nu_\tau$\\
($\pi$ and $K$ decays)&CHORUS~\cite{Eskut:2007rn}, NOMAD~\cite{Astier:2001yj}& &\\
 &MiniBooNE~\cite{AguilarArevalo:2007it}&${\overline \nu_\mu}\not \rightarrow \overline{\nu_e}$\\
\midrule
Accelerator&LSND~\cite{Aguilar:2001ty}&${\overline \nu_\mu}\rightarrow \overline{\nu_e}$\\
($\mu$ decay at rest) &KARMEN~\cite{Armbruster:2002mp}&${\overline \nu_\mu}\not \rightarrow \overline{\nu_e}$\\
\cbottomrule
\end{tabular}
\caption{Neutrino source for the different oscillation experiments and 
search process.} 
\label{tab:NuOscExp}
\end{center}
\end{table}
As we will argue, these seem to be largely 
insensitive to new four-fermion interactions 
involving an electron, a muon and the corresponding neutrinos. 
Reactor experiments are initiated by electron antineutrinos 
from $\beta$ decay, they are then LH and fully described by the SM. 
Similarly, solar neutrinos have electronic flavour, are also 
LH and produced by SM reactions.  
Atmospheric neutrinos are decay products of pions and muons 
from cosmic rays, and may include RH neutrinos.  
However, since the flux of cosmic rays is isotropic, and atmospheric 
neutrino oscillation experiments only compare fluxes of muon neutrinos 
coming from different directions, they 
are not sensitive to a possible deficit in the {\em total} number of initial 
LH neutrinos from muon decays. 
On the other hand, in accelerator experiments
looking for $\nu_\mu \rightarrow \nu_\tau$ or  $\nu_\mu\rightarrow \nu_e$, 
neutrino beams are mainly formed by muon neutrinos originating 
from pion decays (with a fairly small contamination)~\cite{Ahn:2006zza}, 
and then LH and with the SM oscillation pattern. 
Finally, in accelerator experiments looking for 
$\overline\nu_\mu \rightarrow \overline\nu_e$, 
the muon antineutrinos are produced in $\mu ^+$ decays, and
therefore they are sensitive to the NI we are interested in. 
However, what they measure is the number of positrons 
produced by inverse $\beta$ decay, 
looking for an excess of electron anti-neutrinos 
instead of looking for a deficit in the observed number 
of muon anti-neutrinos. 
(The excess reported by the LSND experiment \cite{Aguilar:2001ty} 
has no explanation in this setup, which predicts a deficit.)
Hence, there appears not to be any contradiction between the significant
deficit predicted by the NI considered here and the interpretation
of current oscillation experiments. 

\section{Further phenomenological implications} 
In specific models that contain new fields and interactions there are in
general further observable effects, as for instance the production of
the new particles at large colliders.  This is the case of the simple
model E2 discussed in Section \ref{sec:model}.  If 
this type of NI saturates the EWPD limits,
\be \left| \frac{\lambda_{\mu
e}}{M_\Delta\left[\mathrm{TeV}\right]}\right| \simeq 0.4\ ,
\label{limit}
\ee
implying a relatively large $\lambda_{\mu e}$ and a light $\Delta$.  
The LHC
will be able to uncover such a scalar triplet for $\Delta$ masses up
to $900$ GeV and an integrated luminosity of 30 fb$^{-1}$
\cite{delAguila:2008cj} \footnote{This estimate is larger than those quoted
in \cite{delAguila:2008cj} for in this model the scalar triplet only 
couples to $e$ and $\mu$ but not to taus, which are more difficult 
to identify and have larger backgrounds. 
In contrast with type II see-saw models, 
this triplet is not the only source of 
neutrino masses, and light
neutrino masses and mixings can be reproduced even when
the $\Delta-\tau$ conplings vanish.}.

For this model the relation in Eq.~(\ref{correlation}) translates into
a correlation between the scalar masses $M_{\eta , \Delta}$ and/or the
Yukawa couplings $f_{ee , \mu \mu} , \lambda_{\mu e}$, namely
\be
\Big|\frac{f^\dagger_{ee}f_{\mu\mu}}{M_\eta^2}\Big|^2 \simeq 32\sqrt{2}G_F\
\frac{\lambda^\dagger_{\mu e}\lambda_{\mu e}}{M_\Delta^2}\ .
\label{correlation2}
\ee
This implies that for these NI to have sizable effects at a neutrino
factory we must have relatively large $f_{ee , \mu \mu}$ and light
$\eta$. The main
signals in this case are missing energy plus one or two leptons $\ell
= e, \mu$ because a charged (neutral) $\eta$ has only sizable decays
into $\ell \nu$ ($\nu \nu$) 
\footnote{See \cite{Aaltonen:2008hx} for alternative scalar doublet models.}. 
Thus, $W, WZ$ and $WW$ production can
provide too large irreducible backgrounds for observing these scalars. 
Other SM processes like $t\bar t$ production can also give large 
backgrounds. The search for this scalar doublet is similar to 
left slepton searches assuming that they only decay into a LH charged 
lepton and the lightest supersymmetric particle \cite{delAguila:1990yw}.  
(Although in general sleptons can also have cascade decays and be 
decay products of other supersymmetric particles \cite{Baer:1993ew}.)  
In this case the LHC discovery limit is $\sim 300$ GeV for an integrated 
luminosity of 30 fb$^{-1}$ \cite{Andreev:2004qq},  
but this is assuming that the slepton doublets coupling to the 
first two families are degenerate. However, in our model there is only 
one scalar doublet, scaling the corresponding limit to 250 GeV after 
correcting by the factor of 2. 
It may also happen that though the $\Delta$
and $\eta$ may be too heavy to be directly observed at the LHC
({\it e.g.} $M_\Delta \gtrsim 1$ TeV and $M_\eta \gtrsim 250$ GeV), their
effects can still be observable at a neutrino factory provided $f$ and $\lambda$
fulfill Eqs. (\ref{limit}) and (\ref{correlation2}): $f_{ee , \mu \mu}
\simeq 2\lambda_{\mu e} \gtrsim 0.8$.  Obviously, large enough lepton
colliders are better suited for searching these scalars, since they
couple mainly to leptons, and these colliders allow for a better
kinematical reconstruction.  On the other hand, the relation in
Eq. (\ref{correlation2}), as the cancellation of other possible LF
violating couplings, does not appear to be natural in this simple
model, but they could be in more complicated frameworks.

Finally, we note that one might consider other SM additions/extensions
generating $\delta g_{LL}^V$, as for instance, heavy neutrino singlets
or triplets mixing with the electron or muon.  In theses cases,
however, the relatively
small mixings allowed by EWPD \cite{delAguila:2008pw} do not allow
contributions large enough to produce a sizable deficit.

\section*{Aknowledgements}
We thank J.A. Aguilar-Saavedra, K. Long, S. Pascoli, M. P\' erez-Victoria and 
J. Santiago for discussions.
This work has been partially supported by 
MICINN 
(FPA2006-05294), by Junta de Andaluc\'{\i}a (FQM 101, FQM 437 and FQM 03048),
and by the European Community's Marie-Curie Research Training
Network under contract MRTN-CT-2006-035505 ``Tools and Precision
Calculations for Physics Discoveries at Colliders'', and
and by the U.S. Department of Energy
grant No.~DEFG03-94ER40837;

\newpage

\section*{Appendix}
\label{app:}
Our fits are performed using a $\chi^2$ analysis for the experimental data 
collected in Tables~\ref{tab:ExpDataNZP} and \ref{tab:ExpDataZP}. 
\begin{table}[h]
\begin{center}
{
\begin{tabular}{ c c c } \hline
\ctoprule
Quantity &Experimental Value& Standard Model\\
\midrule
$m_t\left[\mbox{GeV}\right]$&$173.1\pm 1.3$&$173.2$\\
\midrule
$M_W\left[\mbox{GeV}\right]$&$80.399\pm 0.023$&$80.373$\\
\midrule
$g^2_L$&$0.3005\pm 0.0012$&$0.3039 $\\[0.1cm]
$g^2_R$&$0.0311\pm 0.0010$&$0.0301 $\\
$\theta_L$&$2.51\pm 0.033$&$2.46 $\\
$\theta_R$&$4.59\pm 0.41$&$5.18 $\\
\midrule
$g_V^{\nu e}$&$-0.040\pm 0.015$&$-0.0399 $\\
$g_A^{\nu e}$&$-0.507\pm 0.014$&$-0.507 $\\
\midrule
$Q_W\left(\mbox{Cs}\right)$&$-72.74\pm 0.46$&$-73.13$\\
$Q_W\left(\mbox{Tl}\right)$&$-116.4\pm 3.6$&$-116.7$\\
$Q_W\left(e\right)\left(\mbox{M{\o}ller}\right)$&$-0.0403\pm 0.0053$&$-0.0473$\\
\midrule
$\sum_i\left|V_{ui}\right|^2$&$1.0000\pm0.0006$&$1$\\
\midrule
$\sigma^{\nu e\rightarrow \nu \mu}/\sigma^{\nu e\rightarrow \nu \mu}_{\SM}$&$0.981\pm 0.057$&$1$\\
\cbottomrule
\end{tabular}}
\caption{Non Z-pole (pseudo-) observables included in the global fits.} 
\label{tab:ExpDataNZP}
\end{center}
\end{table}
\begin{table}[!]
\begin{center}
{
\begin{tabular}{ c c c } \hline
\ctoprule
Quantity &Experimental Value& Standard Model\\
\midrule
$\Delta \alpha^{(5)}_\mathrm{had}\left(M_Z^2\right)$&$0.02758\pm 0.00035$&$0.02769 $\\[0.1cm]
$\alpha_S\left(M_Z^2\right)$&$0.1184\pm 0.0007$&$0.1184$\\
\midrule
$M_Z\left[\mbox{GeV}\right]$&$91.1876\pm 0.0021$&$91.1875 $\\
$\Gamma_Z \left[\mbox{GeV}\right]$&$2.4952\pm 0.0023$&$2.4957 $\\
$\sigma_H^0\left[\mbox{nb}\right]$&$41.541\pm 0.037$&$41.479 $\\
$R^0_e$&$20.804\pm 0.050$&$20.741 $\\
$R^0_\mu$&$20.785\pm 0.033$&$20.741 $\\
$R^0_\tau$&$20.764\pm 0.045$&$20.788  $\\
$A^{0,e}_\mathrm{FB}$&$0.0145\pm 0.0025$&$0.0164 $\\
$A^{0,\mu}_\mathrm{FB}$&$0.0169\pm 0.0013$&$  $\\
$A^{0,\tau}_\mathrm{FB}$&$0.0188\pm 0.0017$&$  $\\
\midrule
$A_e\left(\mbox{SLD}\right)$&$0.1516\pm 0.0021$&$0.1477 $\\
$A_\mu\left(\mbox{SLD}\right)$&$0.142\pm 0.015$&$  $\\
$A_\tau\left(\mbox{SLD}\right)$&$0.136\pm 0.015$&$  $\\
\midrule
$A_e\left(P_\tau\right)$&$0.1498\pm 0.0049$&$ $\\
$A_\tau\left(P_\tau\right)$&$0.1439\pm 0.0043$&$ $\\
\midrule
$R^0_b$&$0.21629\pm 0.00066$&$0.2158 $\\
$R^0_c$&$0.1721\pm 0.0030$&$0.1722$\\
$A^{0,b}_\mathrm{FB}$&$0.0992\pm 0.0016$&$0.1035  $\\
$A^{0,c}_\mathrm{FB}$&$0.0707\pm 0.0035$&$0.074$\\ 
$A_b$&$0.923\pm 0.020$&$0.935  $\\
$A_c$&$0.670\pm 0.027$&$0.668  $\\
\midrule
$A^{0,s}_\mathrm{FB}$&$0.098\pm 0.011$&$0.1037  $\\
$A_s$&$0.895\pm 0.091$&$0.936  $\\
$R^0_s/R^0_{u+d+s}$&$0.371\pm 0.022$&$0.359 $\\
\midrule
$Q_\mathrm{FB}^\mathrm{had}$&$0.0403\pm 0.0026$&$0.0424 $\\
$\sin^2{\theta_\mathrm{eff}^\mathrm{lept}}$&$0.2315\pm 0.0018$&$0.23143 $\\
\cbottomrule
\end{tabular}}
\caption{Z-pole (pseudo-) observables included in the global fits.} 
\label{tab:ExpDataZP}
\end{center}
\end{table}
This data set is essentially the same used in \cite{delAguila:2008pw}, but 
updating the top \cite{:2009ec} and $W$ \cite{:2009nu} mass values 
as well as the world average for 
$\alpha_s$ \cite{Bethke:2009jm}. 
We also include light flavour data at the Z-pole \cite{:2005ema}, 
the Tevatron determination of the effective weak mixing angle~ 
\cite{Acosta:2004wq}, and further low energy observables, 
such as the weak charges for Thalium (atomic parity violation) and electron 
(M{\o}ller scattering) \cite{Amsler:2008zzb}, the unitarity of the first row of the 
CKM matrix \cite{Hardy:2008gy}, and obviously, the data from inverse muon 
decay \cite{Mishra:1990yf}. The SM best fit values for each observable are also 
included in the Tables. They have been computed using {\tt ZFITTER 6.43} 
\cite{Arbuzov:2005ma}, complemented with routines for the computation of some 
of the low energy observables. For the minimizations of the $\chi^2$ we use the 
package {\tt MINUIT} \cite{James:1975dr}. 
The SM minimum is obtained for the following values:
\bea
m_H=&\hspace{-2cm}96^{+37}_{-28}~\mathrm{GeV},&\hspace{0.8cm}m_t=173.2\pm1.3~\mathrm{ GeV},\nn\\
M_Z=&91.1875\pm0.0021~\mathrm{ GeV},&\alpha_s\left(\scriptmath{M_Z^2}\right)=0.1184\pm0.0007,\nn\\
\Delta \alpha^{(5)}_\mathrm{had}\left(\scriptmath{M_Z^2}\right)=&\hspace{-0.6cm}\left(276.9\pm3.3\right)\cdot 10^{-4}.\ &\nn
\eea
The new physics effects are incorporated adding to the SM predictions 
the tree level contributions from the new operators. 
The impact of this extension on the best fit values for the SM inputs is insignificant. 
For the E1 completion, which allows for the largest departure from the SM, 
\bea
\left.m_H\right|_{E1}=&\hspace{-2cm}92^{+40}_{-29}~\mathrm{GeV},&\hspace{0.8cm}\left.m_t\right|_{E1}=173.2\pm1.3~\mathrm{ GeV},\nn\\
\left.M_Z\right|_{E1}=&91.1875\pm0.0021~\mathrm{ GeV},&\left.\alpha_s\left(\scriptmath{M_Z^2}\right)\right|_{E1}=0.1184\pm0.0007,\nn\\
\left.\Delta \alpha^{(5)}_\mathrm{had}\left(\scriptmath{M_Z^2}\right)\right|_{E1}=&\hspace{-0.6cm}\left(276.8\pm3.3\right)\cdot 10^{-4}.\ &\nn
\eea
Obviously, this is because none of the new operators directly contributes to (almost) 
any of the considered observables; and they indirectly do only through $G_\mu$, 
which is in excellent agreement with the SM prediction. 
Thus, $\delta g_{LL}^V$ tends to be negative in order to compensate the effect of the 
other operators and the SM parameters do not feel the presence of the new parameters.
Once we have introduced the new operators, we perform a scan over the parameter 
space and reconstruct the probability density function (p.d.f.) using an acceptance/rejection 
method. 
From the resulting distribution we compute the confidence regions at 90$\%$ C.L., 
as well as the marginal p.d.f. for each parameter, which we use to obtain the limits and 
confidence intervals quoted in the text.

\newpage



\end{document}